\documentclass[sigconf,screen]{acmart}

\AtBeginDocument{%
  \providecommand\BibTeX{{%
    \normalfont B\kern-0.5em{\scshape i\kern-0.25em b}\kern-0.8em\TeX}}}

\AtBeginDocument{%
	\providecommand\BibTeX{{%
			Bib\TeX}}}

\setcopyright{rightsretained}
\acmDOI{10.1145/3650212.3680357}
\acmYear{2024}
\copyrightyear{2024}
\acmISBN{979-8-4007-0612-7/24/09}
\acmConference[ISSTA '24]{Proceedings of the 33rd ACM SIGSOFT International Symposium on Software Testing and Analysis}{September 16--20, 2024}{Vienna, Austria}
\acmBooktitle{Proceedings of the 33rd ACM SIGSOFT International Symposium on Software Testing and Analysis (ISSTA '24), September 16--20, 2024, Vienna, Austria}
\acmSubmissionID{issta24main-p819-p}
\received{2024-04-12}
\received[accepted]{2024-07-03}

\usepackage[firstpage]{draftwatermark}
\SetWatermarkAngle{0}
\SetWatermarkText{\raisebox{26cm}{%
  \hspace{16cm}%
  \href{https://www.acm.org/publications/policies/artifact-review-and-badging-current}{\includegraphics[width=0.20\linewidth]{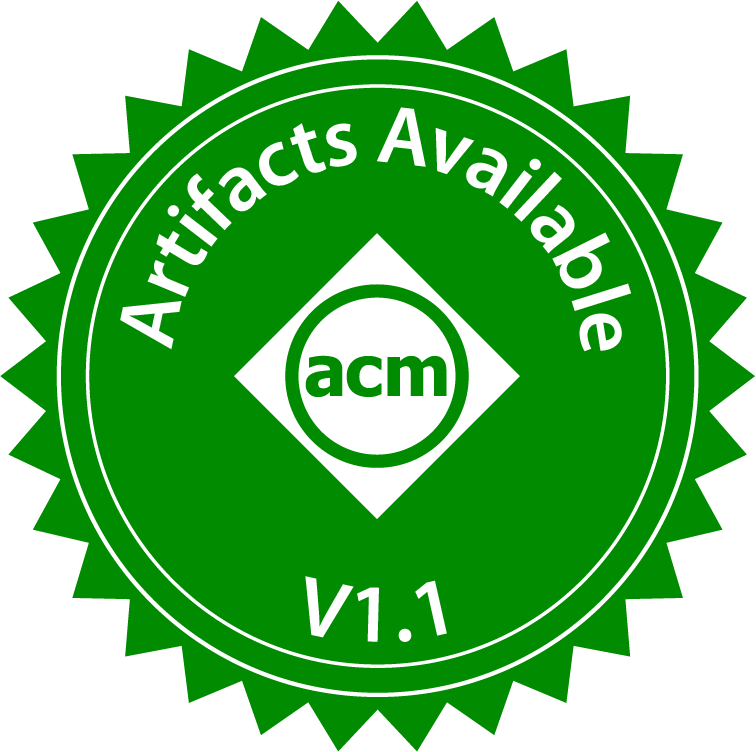}}%
  \href{https://www.acm.org/publications/policies/artifact-review-and-badging-current}{\includegraphics[width=0.20\linewidth]{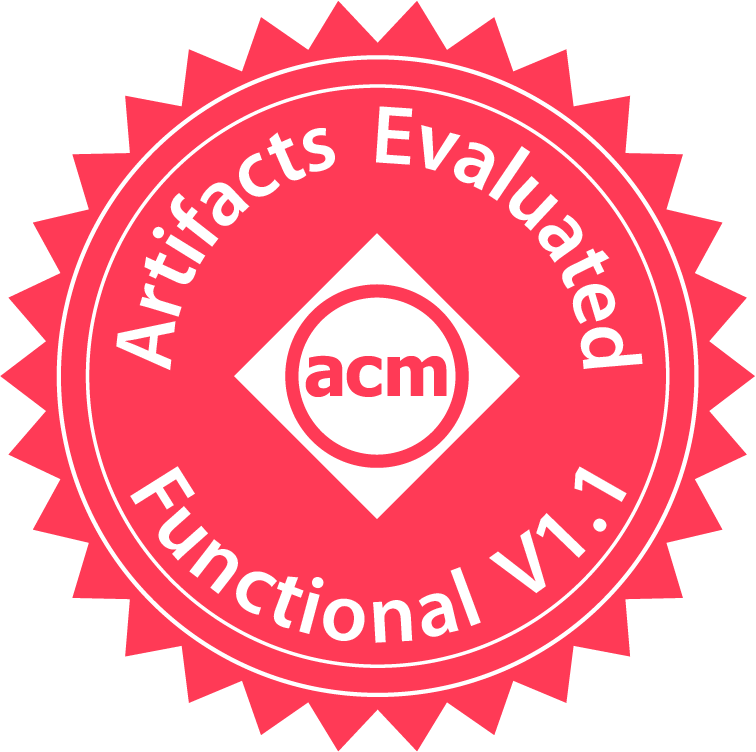}}%
	\href{https://www.acm.org/publications/policies/artifact-review-and-badging-current}{\includegraphics[width=0.20\linewidth]{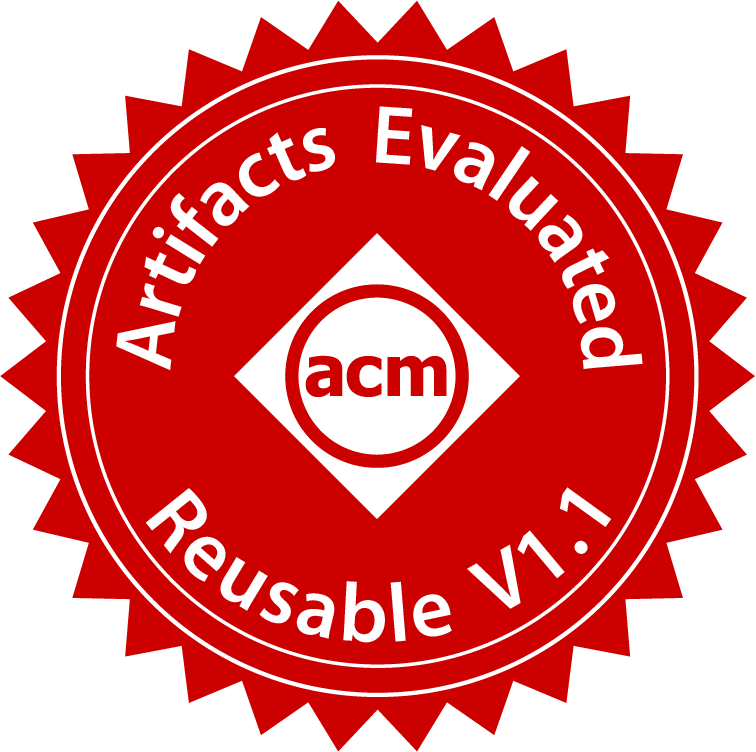}}%
}}

\usepackage[utf8]{inputenc}
\usepackage[T1]{fontenc}
\usepackage{url}
\usepackage{colortbl}
\usepackage{balance}
\usepackage{xspace}
\usepackage{graphicx}
\usepackage{verbatim}
\usepackage{bold-extra}
\usepackage{amsmath}
\usepackage{multirow}
\usepackage{listings}
\usepackage{soul}
\usepackage{enumitem}
\usepackage[normalem]{ulem}
\setlist{nolistsep,leftmargin=.5cm}
\usepackage{booktabs}
\usepackage{tabularx}
\usepackage{svg}
\usepackage{calc}
\usepackage{float}
\usepackage{multirow}
\usepackage[normalem]{ulem}
\useunder{\uline}{\ul}{}
\usepackage{amsmath}
\usepackage{xcolor}
\usepackage{caption}
\usepackage{microtype} 
\usepackage{amsmath,amsfonts}
\usepackage{algorithmic}
\usepackage{graphicx}
\usepackage{textcomp}
\usepackage{hyperref}
\usepackage{ifthen}
\usepackage{wrapfig}
\usepackage{subcaption}
\usepackage{cleveref} 
\usepackage[most]{tcolorbox}

\usepackage{booktabs, multirow} %
\usepackage{soul}%
\usepackage{changepage,threeparttable} %
\usepackage{adjustbox}
\usepackage{hhline}

\newcommand\rev[1]{{\color{black}{#1}}}

{

	\newcommand{\ie}{\textit{i.e.},\xspace}
	\newcommand{\eg}{\textit{e.g.},\xspace}
	\newcommand{\etc}{\textit{etc.}\xspace}
	\newcommand{\etal}{\textit{et al.}\xspace}
	
	\newcommand{\aka}{\textit{a.k.a.}\xspace}

		\usepackage{tikz}

	\newcommand{\crashscope}{{\scshape{CrashScope}}\xspace}

	\newcommand{\lucene}{{\scshape{Lucene}}\xspace}
	\newcommand{\sbert}{{\scshape{SBert}}\xspace}
	\newcommand{\clip}{{\scshape{Clip}}\xspace}
	\newcommand{\blip}{{\scshape{Blip}}\xspace}
        \newcommand{\bugloc}{{\scshape{BugLocator}}\xspace}
	\newcommand{\sentbert}{{\scshape{SentenceBert}}\xspace}
	\newcommand{\bert}{{\sc Bert}\xspace}
	\newcommand{\uibert}{{\sc UIBert}\xspace}
	\newcommand{\vut}{{\sc VuT}\xspace}
	\newcommand{\stov}{{\sc Screen2Vec}\xspace}

	\newcommand{\concatob}{{\textsc{Concat OBs}}\xspace}
	\newcommand{\firstob}{{\textsc{First OB}}\xspace}
	\newcommand{\indivob}{{\textsc{Individual OBs}}\xspace}

	\newcommand{\gator}{{\textsc{Gator}}\xspace}
	\newcommand{\frontmattert}{{\textsc{FrontMatter}}\xspace}
	\newcommand{\backstage}{{\textsc{Backstage}}\xspace}
	\newcommand{\goalexplorer}{{\textsc{GoalExplorer}}\xspace}

	\newcolumntype{M}[1]{>{\centering\arraybackslash}m{#1}}

\begin{document}

\title{Toward the Automated Localization of Buggy Mobile App UIs from Bug Descriptions}

\author{Antu Saha}
\orcid{0009-0004-8656-7581}
\affiliation{%
  \institution{William \& Mary}
  \city{Williamsburg}
  \state{Virginia}
  \country{USA}
}
\email{asaha02@wm.edu}

\author{Yang Song}
\orcid{0000-0002-6490-600X}
\affiliation{%
  \institution{William \& Mary}
  \city{Williamsburg}
  \state{Virginia}
  \country{USA}
}
\email{ysong10@wm.edu}

\author{Junayed Mahmud}
\orcid{0000-0002-5172-1911}
\affiliation{%
  \institution{University of Central Florida}
  \city{Orlando}
  \state{Florida}
  \country{USA}
}
\email{junayed.mahmud@ucf.edu}

\author{Ying Zhou}
\orcid{0000-0002-0949-9755}
\affiliation{%
  \institution{George Mason University}
  \city{Fairfax}
  \state{Virginia}
  \country{USA}
}
\email{yzhou29@gmu.edu}

\author{Kevin Moran}
\orcid{0000-0001-9683-5616}
\affiliation{%
  \institution{University of Central Florida}
  \city{Orlando}
  \state{Florida}
  \country{USA}
}
\email{kpmoran@ucf.edu}

\author{Oscar Chaparro}
\orcid{0000-0003-2838-685X}
\affiliation{%
  \institution{William \& Mary}
  \city{Williamsburg}
  \state{Virginia}
  \country{USA}
}
\email{oscarch@wm.edu}

\begin{abstract}
Bug report management is a costly software maintenance process comprised of several challenging tasks.  
Given the UI-driven nature of mobile apps, bugs typically manifest through the UI, hence the identification of buggy \rev{UI screens and UI components (\textit{Buggy UI Localization})} is important to localizing the buggy behavior and eventually fixing it. \rev{However, this task is challenging as developers must reason about bug descriptions (which are often low-quality), and the visual or code-based representations of UI screens}.
\looseness=-1

This paper is the first to investigate the feasibility of automating the task of Buggy UI Localization through a comprehensive study that evaluates the capabilities of \rev{one textual and two multi-modal deep learning (DL) techniques and one textual unsupervised technique.} 
We evaluate such techniques at two levels of granularity, \rev{Buggy \textit{UI Screen} and \textit{UI Component} localization}. Our results illustrate the individual strengths of models that make use of different representations, wherein models that incorporate visual information perform better for \rev{UI screen localization}, and models that operate on textual screen information perform better for \rev{UI component localization} -- highlighting the need for a localization approach that blends the benefits of both types of techniques. 
Furthermore, we study whether Buggy UI Localization can improve traditional \rev{buggy code localization}, 
and find that incorporating localized buggy UIs leads to \rev{improvements of 9\%-12\% in Hits@10}.

\end{abstract}



\maketitle

\section{Introduction}
\label{sec:intro}

Bug report management is an essential, yet costly process for software projects, in particular for mobile apps~\cite{zou2018practitioners}. It demands high developer effort~\cite{zou2018practitioners,Zimmermann2010,Anvik2006,fan2018chaff,Bissyande2013,Tian2015a} due in part to the potential for large volumes of reported bugs and the varying quality of submitted bug reports. These reports are the central artifact in bug management~\cite{Zimmermann2010,zou2018practitioners,fan2018chaff,Chaparro2017-2}, as they directly impact downstream tasks such as bug triaging, reproduction, localization, program repair, and even regression testing. Bug reports typically describe defects found during software development and usage, and are expected to include, at minimum, the app’s \rev{observed (incorrect) behavior~(\textbf{OB}}, the expected behavior (\textbf{EB}), and the steps to reproduce the bug (\textbf{S2Rs})~\cite{Bettenburg2008a,Laukkanen2011,Zimmermann2010,Chaparro2017-2,soltani2020significance,song2020bee}.

\rev{
Given the UI-centric nature of mobile apps, a large majority of reported bugs for these apps manifest through the UI~\cite{johnson2022empirical}. Therefore, an important first step toward understanding, diagnosing, and resolving underlying bugs in the code is localizing the buggy behavior to both a UI screen and UI components~\cite{mahmud2023using}. 
As such, a critical bug report management task for mobile apps is the identification of \rev{UI screens and UI components (\eg buttons or text fields)} that cause or display the reported incorrect behavior of the app (\ie the OB), a task that we term \textbf{\textit{Buggy UI Localization}}.
}
This task is essential but can be difficult for developers, especially when many incoming bug reports need to be addressed and fail to include important details or graphical information (\eg buggy app screenshots~\cite{erfani2014works}). 
Despite the growing body of work on \rev{automating bug report management tasks~\cite{zou2018practitioners},} prior work has not yet explored how to assist developers in Buggy UI Localization.
\looseness=-1

\looseness=-1

\looseness=-1 

In this paper, we present the first empirical study that investigates \textit{the feasibility of automatically localizing bug descriptions to \rev{UI screens and UI components} of mobile apps}. Similar to traditional \rev{buggy code localization}~\cite{akbar2020large,
Wong2014,Kochhar2014,florez2021combining,chaparro2019using,Chaparro2017-3,chaparro2019reformulating}, we formulate Buggy UI Localization as a retrieval task, in which a bug description \rev{(\ie the \textbf{OB})} is used as query input to a retrieval engine that searches the space of \rev{UI screens and UI components} of an app and recommends a ranked list of candidates that most likely correspond to the bug description.
\rev{Specifically, the study focuses on two retrieval tasks for a given bug description: \textbf{\textit{screen localization (SL)}}, which involves retrieving potentially buggy UI screens from the app, and \textbf{\textit{component localization (CL)}}, which aims to retrieve the relevant buggy UI components from a given buggy UI screen.
}

\looseness=-1

The study investigates how the textual and visual information from \rev{UI screens and UI components} can be leveraged for Buggy UI Localization, and hence, explores the effectiveness of \rev{unsupervised textual techniques and} pre-trained textual and multi-modal deep learning (DL) techniques. Specifically, we examine \rev{one unsupervised text-based model (\lucene~\cite{Hatcher2004}) and} three DL models: a supervised text-based model (\sentbert or  \sbert~\cite{reimers2019sentence}), and two supervised vision-language learning models (\clip~\cite{radford2021learning} and \blip~\cite{li2022blip}), under a zero-shot setting to explore their capabilities for Buggy UI Localization. To evaluate the effectiveness of the models in real-world scenarios, we created a manually curated dataset of 228 OB descriptions from 87 real bug reports. The dataset also includes associated buggy \rev{UI screens and UI components} that we manually labeled from a UI corpus created by employing GUI app exploration techniques~\cite{Moran2016}, for 39 Android mobile apps.
\looseness=-1

The results of our study indicate that no single technique universally performs best for the two localization tasks \rev{(screen and component localization)}. The best-performing approaches suggest the correct buggy UI screens (\blip) and \rev{UI components} (\sbert) in the top-3 recommendations for 52\% and 60\% of the bug descriptions, respectively. We also found the models tend to perform better for higher-quality bug descriptions and easier-to-retrieve cases as judged by humans. The results show the feasibility and effectiveness of using existing DL models for \rev{Buggy UI Localization}.
\looseness=-1

To illustrate the practical usefulness of automated Buggy UI Localization, we \rev{conducted a second empirical study that investigated} how identified buggy UI screens from the best-performing \rev{screen localization} model can improve \rev{traditional buggy code localization approaches}.
We adapted the approach proposed by Mahmud \etal~\cite{mahmud2023using} to filter or boost code files retrieved via existing \rev{buggy code localization approaches using retrieved UI screen information by screen localization.}
\rev{
We designed an end-to-end, automated approach comprising two major steps: (1) \textbf{\textit{Buggy UI Localization}}, which receives a bug description and the app UI screens 
and automatically identifies buggy UI screens, and (2) \textbf{\textit{Buggy Code Localization}}, which (i) computes the textual similarity between the bug report and the app code files to retrieve potentially buggy files, and (ii) boosts the rankings of retrieved code files related to identified buggy UI screens in step (1).}
Using two \rev{buggy code localization} tools applied to 79 bug reports we found that incorporating information from the automatically identified 
Buggy UI screens can lead to a \rev{9\% to 12\% improvement in 
Hits@10}, compared to baseline techniques that do not use UI data.
\looseness=-1

\rev{In summary, we make the following contributions:
\begin{enumerate}
    \item We are the first to address the problem of \textbf{Buggy UI Localization} 
   by investigating how different information sources (textual UI metadata and app screenshots) and  four existing textual and multimodal models can be leveraged for Buggy UI Localization. 
    Our results suggest that the models perform differently depending on the retrieval task. These findings can inform the design of future domain-specific models.
    
    \item We illustrate a practical application of buggy UI screen localization on  \textbf{Buggy Code Localization}. The application of our screen localization approach to Mahmud \etal's approach~\cite{mahmud2023using} illustrates that it can both automate and improve upon existing buggy code localization techniques. 
    
    \item We provide a novel, publicly available benchmark (data, infrastructure, results, and documentation) for Buggy UI Localization, which facilitates replication and experimentation~\cite{package}. The benchmark provides a new, manually-curated dataset with buggy UI screens and UI components, textual and visual retrieval corpora, and bug descriptions for each bug report.
    \looseness=-1
\end{enumerate}

}

\begin{figure}[t]
	\centering

	\begin{tcolorbox}[left=2pt,right=2pt,top=0pt,bottom=0pt]
		\small
		\textbf{Title:} \underline{Can no longer enter text in SSID Filter TextView }\\
		
		\textbf{Description:}  \underline{Cannot enter any text in the SSID Filter field.}\\
		Steps:
		
		1. Click on Filter icon.\\
		2. Click/tap on SSID Filter text field.\\
		3. \underline{Keyboard does not pop up.}\\
		
		Expected Behaviour:
		
		Should display keyboard and allow you to enter SSID filter text.
	\end{tcolorbox}
	\caption{Bug report \#191 from the WiFi Analyzer app~\cite{wifianalyzerbug} 
	}
	\label{fig:bug-example}
\end{figure}

\section{Background, Problem, \& Motivation}
\label{sec:problem_motivation}

\subsection{Bug Descriptions \& App UI Elements}

In this paper, a \textit{\textbf{bug description}} is the observed or incorrect app behavior (\textbf{OB}) textually described in a sentence of a bug report. We focus on descriptions of bugs that manifest visually on the device screen.
\Cref{fig:bug-example} shows 
a real bug report for WiFi Analyzer~\cite{wifianalyzerbug}, an app for monitoring the strength and channels of surrounding WiFi networks~\cite{wifianalyzer}. The bug/OB descriptions in the bug report are underlined \rev{in \Cref{fig:bug-example}} and describe a bug in which the app fails to show the keyboard to enter the WiFi's SSID.  

\begin{figure*}[t]
	\centering
	\includegraphics[width=\linewidth]{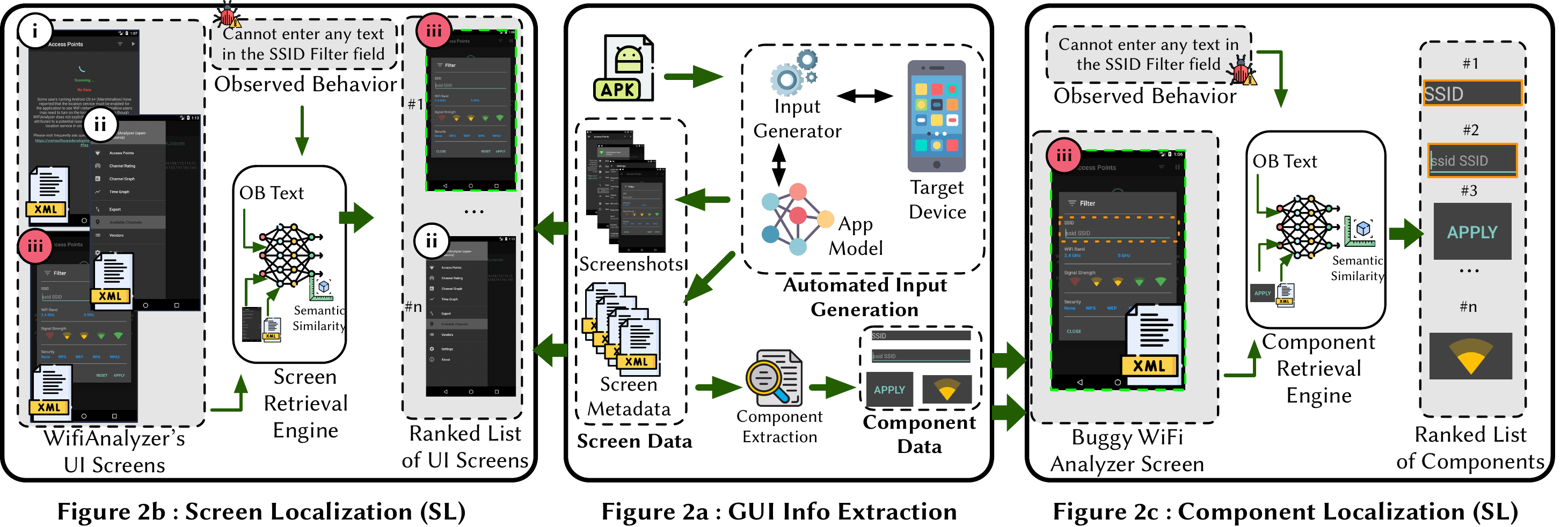}
	\caption{Example of the UI screen/component localization process for an OB/bug description of the WifiAnalyzer app~\cite{wifianalyzerbug}.}
	\label{fig:retrieval_tasks}
\end{figure*}

App \textit{\textbf{UI screens}} implement one or more app features and represent the canvas upon which UI components (\aka widgets) are drawn. \textit{\textbf{UI components}} are elements rendered on a UI screen (\eg buttons, text fields, or checkboxes) that allow end-users to interact with the application. A screen is composed of a hierarchy of UI components and containers (\aka layouts) that group UI components together~\cite{AndroidLayout}. \rev{\Cref{fig:retrieval_tasks} shows examples of UI screens (2b) and their components (2c) for the  WiFi Analyzer app.} 
In this paper, a UI screen is represented as a \textit{screenshot} and its corresponding \textit{UI hierarchy} of components/containers described in metadata. Each UI component is represented by a set of \textit{attributes}, including the component type (\eg TextView or Button~\cite{AndroidLayout}), its label or text shown on the screen, an ID, a description, and various visual properties such as the component's visibility and size. \rev{\textbf{Buggy UI Screens and UI Components}} display unexpected, incorrect behavior of an app.
\looseness=-1

\subsection{Problem and Motivating Example}

We envision a system that suggests to the developer a ranked list of UI screens (\ie app screenshots) that display or is related to \rev{the buggy app behavior reported by a bug description in a bug report (see \Cref{fig:retrieval_tasks}b).}
The developer would then inspect the suggested UI screens \rev{in the ranked list (in a top-down fashion)} and select one or more screens that s/he deems display the reported bug. The system would then identify (and highlight) the \rev{UI components} in the selected buggy UI screens that are most related to the reported bug \rev{(see \Cref{fig:retrieval_tasks}c)}. The suggestions of this system can help developers not only automatically localize \rev{buggy UI screens and UI components}~\cite{nayebi2020eye,yusop2016reporting,agrawal2022understanding,Breu2010,Zimmermann2010}, but also understand the reported bug, and assist them in other bug management activities (\eg bug reproduction). Additionally, this system can be useful for various bug report management tasks, as it can provide information to existing automated techniques that aim to reproduce bugs~\cite{Zhao2019,zhao2022recdroid+}, generate test cases~\cite{Fazzini2018}, assess the quality of bug reports~\cite{song_toward_2022,Chaparro:FSE19,song_burt_2023}, and perform \rev{buggy code localization}~\cite{mahmud2023using}. 
\looseness=-1

While bug reports provide the steps to reproduce the bug (S2Rs) \rev{and the expected app behavior (EB)}, which can be used to identify the \rev{buggy UI screens and UI components}, we focus on OB descriptions for at least two reasons~\cite{Chaparro2017-2}: (1) they convey the faults observed by the user, and (2) they are often written using different wordings (even for a single bug type---see \cref{fig:bug-example}). The S2Rs \rev{and EB} in bug reports do not necessarily describe a bug \rev{and they are often described using a more limited language compared to that of OBs}~\cite{Chaparro2017-2}.
\looseness=-1

We formulate automated buggy UI localization as two retrieval tasks \rev{(see \Cref{fig:retrieval_tasks})}: \textit{screen} and \textit{component} localization.   In \textit{\textbf{screen localization (SL)}}, a bug/OB description (\ie the~\textit{\textbf{query}}) is the input to a retrieval engine that searches the space of \rev{(automatically identified) UI screens (see \Cref{fig:retrieval_tasks}a)} of a given app and retrieves a list of UI screens ranked by their similarity to the bug description, which indicates the likelihood of a UI screen to show or be affected by the bug described by the query. 
\rev{\Cref{fig:retrieval_tasks}b illustrates} the screen localization process for one OB description from the bug report shown in \cref{fig:bug-example}. The highlighted UI screen with the green border \rev{(see Screen (iii) of \Cref{fig:retrieval_tasks}b)} is the buggy screen (initially unknown to the developer).
The two best approaches we studied (\blip \& \sbert) are able to retrieve the buggy screen as their first suggestion. 
In \textbf{\textit{component localization (CL)}}, the retrieval engine searches the space of \rev{(automatically identified) UI components (see \Cref{fig:retrieval_tasks}a) of a given buggy UI screen} and retrieves a list of \rev{UI components} ranked by a similarity score that indicates the likelihood of the components to show or be affected by the bug. \rev{\Cref{fig:retrieval_tasks}c illustrates} the component localization process for the buggy UI screen of the bug description. The  \rev{UI components} in orange are the ones that the bug description refers to, hence they are expected to be ranked higher by the component retrieval engine. 
The two best-performing approaches we studied (\sbert \& \blip), rank the buggy components in the first position(s). %
\looseness=-1

Screen and component localization are impacted by the amount of information that a OB description contains (\ie \textbf{\textit{query quality}}) and the difficulty in retrieving \rev{buggy UI screens/UI components} (\ie \textbf{\textit{retrieval difficulty}}). If the bug description is poorly written or does not provide enough information about the bug (which is  not uncommon in bug reports~\cite{Chaparro:FSE19,song_toward_2022}), then a retrieval engine (or even a human) would have a hard time identifying the \rev{buggy UI screens and UI components} (if not familiar enough with the app). This problem is exacerbated by the fact that the same bug can be described in a variety of ways~\cite{Chaparro2017-2} -- \rev{\eg see the underlined sentences in \cref{fig:bug-example}}. Even if the OB is clear and informative, identifying the \rev{buggy UI screens/UI components} can be challenging when numerous similar \rev{UI screens/UI components} exist in the app.
As an example, consider the last OB/bug description from the Wifi Analyzer app shown in \cref{fig:bug-example}: ``Keyboard does not pop up". 
\rev{The best approach for screen localization, \blip, retrieved the true buggy screen at the 21st position and true buggy component at the 16th position. Component localization's best approach, \sbert, retrieved the true buggy screen at the 6th position and the true buggy components at the 12th position.}
This illustrates the difficulty of buggy UI localization, hence in \rev{our} study, we assess the performance of various approaches considering bug descriptions of different quality and retrieval difficulty levels.
\looseness=-1

\section{Automating Buggy UI Localization}
\label{sec:methodology}

This study aims to investigate different methods for automatically locating \rev{buggy UI screens and UI components} based on bug/OB descriptions and measure their effectiveness for this problem. 
To that end, we investigate existing retrieval approaches that leverage textual and/or visual information from the bug descriptions and \rev{UI screens and components}, to perform screen and component localization.  With this in mind, we address three research questions (RQs):
\looseness=-1

\begin{enumerate}[label=\textbf{RQ$_\arabic*$:}, ref=\textbf{RQ$_\arabic*$}, itemindent=0cm,leftmargin=1cm]
    \item \label{rq:screen_retr}{\textit{How effective are retrieval approaches at locating buggy UI screens (SL) from bug descriptions?}}
    \item \label{rq:component_retr}{\textit{How effective are retrieval approaches at locating buggy UI components (CL) from bug descriptions?}}
    \item \label{rq:query_quality}{\textit{How effective are retrieval approaches for different query quality and retrieval difficulty levels?}}
\end{enumerate}

To answer the RQs, we selected \rev{four (un)supervised approaches} of various kinds (\cref{subsec:approaches}). Then, we constructed a real-world dataset for evaluating the effectiveness of the approaches (\cref{subsec:real_data}). We executed the approaches (\cref{subsec:model_execution}) and measured their performance with standard retrieval metrics (\cref{subsec:metrics}). 
\rev{This section provides details about these steps, while \cref{sec:results,sec:results_discussion} present and discuss the obtained results.}
\looseness=-1

\looseness=-1

\subsection{Retrieval Approaches}
\label{subsec:approaches}

We investigated \rev{three deep learning (DL)-based approaches and one baseline unsupervised approach}, which support text-to-text or text-to-image retrieval for the two Buggy UI Localization tasks.

\textbf{\sentbert} (or \textbf{\sbert})~\cite{reimers2019sentence} is a neural text-based language model, which augments the traditional \bert model~\cite{devlin2018bert} with siamese and triplet networks.
It can establish semantic similarity for a pair of textual descriptions by generating embeddings.
\sbert can be utilized for both screen and component localization using the textual bug descriptions, \rev{UI screens, and UI components}.

\textbf{\clip}~\cite{radford2021learning} is a neural multi-modal vision/language model that can learn semantic embeddings from text and images via a contrastive architecture. Given a text-image pair, it can determine the similarity between them. 
Hence, \clip can be utilized for both screen and component localization using textual information from the bug description and visual information from \rev{UI screen/components}.
\looseness=-1

\textbf{\blip}~\cite{li2022blip} excels in vision-language understanding and generation tasks, and uses a multi-modal encoder-decoder component (MED) and a dataset bootstrapping method, \ie captioning-filtering (CapFilt). 
We used a \blip version optimized for text-image retrieval tasks, which only implements contrastive and matching losses.  \blip can be utilized the same way \clip is used for buggy UI localization.
\looseness=-1

Finally, we selected \textbf{\lucene}~\cite{Hatcher2004} as a baseline technique for text retrieval.  \lucene is a classical unsupervised approach that combines the vector space model (VSM), based on the TF-IDF representation, and the boolean text retrieval model, to compute the (cosine) similarity between a query and a document. \lucene can be utilized for both screen and component localization using the textual information in bug descriptions, \rev{UI screens, and UI components}.
\looseness=-1

While \clip and \blip have been pre-trained with general-purpose data, they 
have performed well under zero-shot settings~\cite{reimers2019sentence,li2022blip,radford2021learning} for tasks such as semantic similarity computation, object detection, image captioning, and text-image retrieval, under distinct domains. These models have also been fine-tuned for downstream tasks~\cite{zhang2022pointclip,baldrati2022conditioned,conde2021clip,rasheed2022fine}, enhancing their capabilities. 
This study investigates the capabilities of these models for Buggy UI Localization. 
\looseness=-1

The lack of a large, high-quality dataset specifically created for Buggy UI Localization prevents us from fine-tuning \clip and \blip. While the RICO dataset~\cite{rico} would be a good dataset candidate for our task, 
it does not provide bug descriptions, and the mobile app screenshots included in RICO do not show any buggy behavior. As such, creating this dataset with real-life bug descriptions and buggy UI screens of mobile apps would demand an enormous effort that is beyond the scope of our work. A potential solution to create this dataset, which we leave for our future work, is creating synthetic bug descriptions using templates for different bug types and wordings, based on automatically modified RICO screenshots that show various buggy behaviors (\eg incorrect app output, crashes, non-crashing errors, cosmetic issues, and navigation misbehavior).
\looseness=-1

Besides the three DL models, we considered models specifically designed for mobile app UI understanding tasks, including \uibert~\cite{bai2021uibert}, \vut~\cite{li2021vut}, and \stov~\cite{li2021screen2vec}. However, \uibert and \vut's source code and pre-trained models are not available and \stov would require a significant adaptation effort for our task as the model is only designed for generating UI screen embeddings from screen text and UI hierarchies; extra modules would be required to adapt this model for Buggy UI Localization. We should note, though, that we experimented with it as a zero-shot encoder to represent UIs and with an \sbert model for representing bug descriptions, computing the cosine similarity on both embeddings to establish similarity. Unfortunately, this led to poor performance for both \rev{screen and component localization tasks}, hence we decided to not report their performance in this paper. 
\looseness=-1

\rev{Large language models (LLMs), \eg  ChatGPT~\cite{openai2023chatgpt} and Llama~\cite{llama2023}, can be used for text-based Buggy UI Localization. However, studying their capabilities for this task likely warrants a separate study because it requires careful control of several factors to make for a fair comparison with non-LLMs, including addressing non-deterministic responses, possible data leakages, selection of bug reports, token limits, and prompting strategies. Studying the capabilities of LLMs for the task of Buggy UI Localization is part of our future work.
}

\subsection{Dataset Construction}
\label{subsec:real_data}
We built a dataset of real-life bug descriptions and relevant \rev{buggy UI screens and UI components} to assess the effectiveness of the models in a realistic setting. \rev{The dataset construction process included: 
\begin{enumerate}
	\item identifying bug/OB descriptions in a set of bug reports -- these descriptions represent the \textbf{queries} used for retrieval.
	\item building the \textbf{UI screen corpus} and  \textbf{the UI component corpus} used for retrieval. These corpora are constructed for the app corresponding to each bug report.
	\item identifying the \textbf{(ground truth) buggy UI screens and UI components} in the corresponding corpus, and assigning a quality and retrieval difficulty level to each query.
\end{enumerate}
}

\looseness=-1

\subsubsection{\textbf{Bug Report Selection}}
\label{subsec:bug_report_selection}

Since one of our goals, later detailed in this paper (see \cref{sec:bug_loc_study}), is to assess the usefulness of Buggy UI Localization models for Buggy Code Localization, 
we followed a pragmatic approach to select the bug reports for this study. \rev{We started with the 80 bug reports of the buggy code localization dataset provided by Mahmud \etal~\cite{mahmud2023using} so that we could reuse the data for this study and the buggy code localization study reported in \cref{sec:bug_loc_study}.}
\rev{Mahmud} \etal's dataset was created based on the AndroR2 dataset~\cite{wendland2021andror2, Johnson2022AndroR2+}, which consists of 180 manually reproduced bug reports for popular open-source Android applications hosted on GitHub. 
\rev{Of the 180 bug reports, Mahmud \etal discarded 100 reports because they: (1) described bugs that were no longer reproducible, (2) included bug fixes in non-Java code files only, (3)  were no longer publicly available, or (4) included ambiguous code changes or commit IDs. This resulted in 80 bug reports.} 
\looseness=-1

\rev{When collecting ground truth data for Buggy UI Localization (see \cref{subsec:ground_trth_mapping}), we discarded one bug report (from the GnuCash app~\cite{gnucash_bug2}) from the 80 bug reports of Mahmud \etal's dataset} because we were unable to reproduce the reported bug, thus leaving 79 reports. To expand the set of bugs usable for this study, we selected 14 extra bug reports from the 100 discarded ones whose bug fixes were in XML resource files as opposed to Java code and discarded six reports because we obtained errors trying to collect the retrieval corpus for those reports (see \cref{subsec:retrieval_corpus}). This resulted in eight extra bug reports, for a total of 87 reports. 
\rev{Although Mahmoud \etal~\cite{mahmud2023using} could not use these 8 bug reports, they are usable for the buggy UI localization task since they are reproducible and we collected (ground truth) UI data for the corresponding applications.}

From the \textbf{87 bug reports} (1 to 8 per app), 32 describe an output problem, 23 report an app crash,  23 describe a UI cosmetic issue, and 9 report a navigation problem. The bug reports correspond to 39 Android apps (\eg GnuCash~\cite{GNU}, Mozilla Focus~\cite{bugzilla}, K-9 Mail~\cite{K-9}, WiFi Analyzer~\cite{wifianalyzerbug}, Images to PDF~\cite{Images2PDF}) of different domains (\eg finance tracking, web browsing,  emailing, WiFi network diagnosis, and image conversion) and UI layouts.

\subsubsection{\textbf{Bug Description Identification}}

\rev{To identify bug/OB descriptions in the 87 bug reports}, two authors inspected and annotated the 1807 sentences in the reports' title and description.  
\rev{Based on the definition of OB and the criteria to annotate OB sentences defined by Chaparro \etal~\cite{Chaparro2017-2}, one author labeled each bug report sentence as either an OB or non-OB. 
Here, the OB sentences describe the buggy app behavior (\eg the underlined sentences in \Cref{fig:bug-example}). 
The second author then verified the annotations made by the first author, indicating agreement or disagreement.
Out of the 1807 sentences across all bug reports, the authors reached agreement on the labels for 1774 sentences ($\approx$98\% agreement, 0.91 Cohen's kappa \cite{Cohen}).}
The authors solved disagreements via discussion and consensus. Reasons for disagreement included mostly mistakes and misinterpretations (\eg when sentences described root causes in the code, rather than UI faults). 
\rev{Finally, 228 sentences were identified as bug/OB descriptions for the 87 bug reports (2 or 3 OBs per bug report on average), which serve as queries for the UI screen and component localization tasks.}

\subsubsection{\textbf{Retrieval Corpus Collection}}
\label{subsec:retrieval_corpus}

To build the retrieval corpus for each bug description, we require the set of \rev{UI screens and UI components} of the apps, including the \rev{buggy UI screens and UI components}. To collect these data, we employed \rev{a semi-automatic app execution approach that consisted of}: (1) a record-and-replay methodology (used in prior studies~\cite{Cooper:ICSE21,song_toward_2022}), and (2) an automated app exploration methodology (used by Chaparro \etal~\cite{Chaparro:FSE19}). \rev{Both  methodologies reduce the manual effort of collecting UI (meta)data.}
\looseness=-1

The goal of the \rev{\textbf{record-and-replay methodology}} was to collect the buggy UI screens for each bug report and the screens navigated while reproducing the bugs, \rev{including screenshots and related metadata}. Two authors manually reproduced the reported bugs by executing the reproduction steps found in the bug report on a Pixel~2 Android emulator. 
While reproducing the bugs, the authors used the AVT tool~\cite{Cooper:ICSE21,song_toward_2022} to collect UI-event traces and a video showing the user interactions with the app and the bug itself \cite{Moran:2017, Moran2016}. These traces were replayed on the emulator via the TraceReplayer tool~\cite{mahmud2023using} to automatically collect app screenshots, UI hierarchies, and metadata for the exercised app UI screens.
\looseness=-1

The goal of the \rev{\textbf{automated app exploration}} was to collect as many UI screens as possible for building the corpus. We executed a version of the \crashscope tool~\cite{Moran:2017} that implements multiple exploration strategies to interact with the UI components of app screens comprehensively, trying to exercise as many app screens as possible. In the process, \crashscope collects app screenshots and XML-based UI hierarchies/metadata for the exercised app UI screens, in the same manner as TraceReplayer.

Since these two methodologies can generate duplicate UI screens, we employed the approach by Chaparro~\etal~\cite{Chaparro:FSE19} to produce a unique set of UI screens for each of the 87 bug reports. This approach parses the UI hierarchies of the collected UI screens for an app and establishes the uniqueness between two screens: \rev{if they have the same hierarchical structure, based on component types, sizes, and parent-children relationships, they are considered the same screen and one of the two is used.} This implies that two UI screens with the same structure but different textual information are considered the same UI screen. 
To create the UI component \rev{corpus for a given (buggy) UI screen}, we parsed the UI hierarchy of the screen and identified the visible \rev{leaf components}, which are typically the ones shown to the user on the mobile device. However, we discarded layouts and other containers, thus focusing on labels, buttons, text fields, and other UI components that users typically interact with.

This procedure resulted in UI screen corpora containing $\approx$26 UI screens per bug report on average,  which \rev{are used} for screen localization.  The UI component corpora contain $\approx$17 UI components per (buggy) UI screen on average, which \rev{is used} for component localization.
\rev{A potential limitation of our corpus collection process, based on dynamic app exploration, is the possibility of missing UI screens for an app, which may affect models' performance. 
However, we evaluate every approach using the same collected retrieval corpus for each bug report, ensuring a fair evaluation. We discuss the (dis)advantages  of a static analysis-based approach for corpus collection over dynamic analysis approaches in \Cref{sec:related_work}.

}

\subsubsection{\textbf{Ground Truth Construction}} 
\label{subsec:ground_trth_mapping}

\rev{We used a rigorous data annotation procedure to identify the buggy UI screens for each bug report, and the buggy UI components for each buggy screen.}

During multiple annotation sessions, four paper authors (\aka annotators) first read and understood the reported bugs, watching (if needed) the bug reproduction video collected during the corpus collection step.  
Then, the annotators inspected the app screens from the corpus to identify the buggy screens shown in the video and marked them as such in a spreadsheet. 
The annotators identified and marked the buggy UI components in the same spreadsheet. Each bug report was assigned to two annotators, making sure the annotators had an even number of bug reports to annotate. For each bug report, the first annotator identified the \rev{buggy UI screen and UI components} and then the second annotator validated \rev{whether the identified screens and components were indeed buggy}. Both annotators \rev{followed} the procedure described above, marking potential disagreements in a shared spreadsheet. At the end of each annotation session, the annotators discussed disagreements (mostly due to misinterpretation of the bugs), and \rev{reached} a consensus to produce the final \rev{ground truth set of buggy UI screens and UI components}.

Besides identifying the \rev{buggy UI screens and UI components}, the annotators rated the quality of the bug/OB descriptions based on the amount of information they provided to understand the bug.
\rev{Since Buggy UI Localization is performed using individual bug/OB descriptions, the annotators judged the quality of each OB in a bug report independently. The detailed understanding of each bug report and the identified buggy UI screens and components assisted the annotators in assessing OB quality.}
The annotators agreed on a quality rating based on a 1-5 discrete scale. A rating of 1 means the bug description does not contain useful information to understand the problem. Conversely, a rating of 5 means the description contains complete information to understand the bug. A rating between 1 and 5 indicates that there is missing information in the OB that hinders bug comprehension. 
Additionally, the annotators marked each bug description as \textit{easy} or \textit{hard} to localize, based on the difficulty they encountered in identifying the \rev{buggy UI screens and UI components}. A common reason why bug descriptions were judged as \textit{hard to retrieve} was
that multiple \rev{UI screens (and UI components)} were similar, yet only one or a few were displaying the reported bug. During the reconciliation sessions, disagreements were discussed and solved to produce the final query quality and retrieval difficulty category for each bug description.

\subsubsection{\textbf{Summary of the Collected Retrieval Data}} 
For screen localization (SL), \rev{each OB description (\ie the query) represents a unique UI screen retrieval task. Hence, our dataset contains 228 queries in total,  with 2.1 buggy UI screens per query as ground truth and 26 UI screens in the corpus on average (see \cref{tab:real_data_stats}).}

For component localization (CL), \rev{each OB description can have multiple ground truth buggy UI screens, hence each combination of OB description and UI screen represents a single retrieval task.  Based on this, we created 254 queries (or retrieval tasks), with 1.9  buggy UI components per buggy UI screen as ground truth, and 17 components in the corpus on average.}
\looseness=-1

In summary, we collected: OB descriptions (\ie the queries), the retrieval corpora of \rev{UI screens and UI components} for each query (including app UI screenshots and cropped component images, and their UI hierarchy with associated metadata: component text, ID, \etc), and ground truth \rev{buggy UI screens and UI components}.

\begin{table}[t]
\centering
\caption{Screen and component localization statistics}
\label{tab:real_data_stats}
\setlength{\tabcolsep}{2pt}
	
\begin{adjustbox}{width=\columnwidth}
\begin{tabular}{c|c|c}
\toprule
\textbf{Statistic}                        & \textbf{Screen loc.} & \textbf{Component loc.} \\ \hline
{\# of retrieval tasks/queries}                  & 228         & 254         \\ \hline
{\# of hard-to-retrieve tasks}     & 111         & 130         \\ \hline
{\# of  easy-to-retrieve tasks}     & 117         & 124         \\ \hline
{Avg. \# of buggy UI screens/comp.}                      & 2.06 (2)    & 1.86 (1)    \\ \hline
{Avg. of corpus size}                   & 25.97 (22)  & 17.11 (14)  \\ \bottomrule
\multicolumn{3}{c}{\scriptsize Average (median) values per query/retrieval task}  \\
\end{tabular}
\end{adjustbox}
\vspace{-2em}
\end{table}

\subsection{ Execution of the Retrieval Approaches}
\label{subsec:model_execution}

\rev{Each retrieval approach processes the query and retrieval corpus differently. Some approaches rely solely on textual information, while others utilize both textual and visual information.}

\clip and \blip leverage textual and visual information from the query and \rev{UI screens and components}. The query for \rev{screen and component localization (SL \& CL)} is \rev{the text of a OB description}. For SL, the corpus is all the screenshots of the application for which the bug is reported, while for CL, the corpus is all the cropped \rev{UI component images of a buggy UI screen}. The models receive a text-image pair and produce a  score indicating how similar the bug/OB description and each \rev{UI screen and UI component} are. 

\lucene and \sbert leverage only the textual information from the query and \rev{UI screens and UI components}. The query is the \rev{OB descriptions}. As for the corpus, we extracted and concatenated the text found in UI component metadata \rev{(\ie the component ID, label, and type)} to create textual documents for retrieval. For CL, each document is represented by the extracted document for a component. For SL, we concatenated the textual documents of the components in a given screen to form the textual document of a screen. \lucene and \sbert compute a score that represents how similar the bug description and each textual document are. Only for \lucene, we applied standard textual pre-processing on the queries and documents (lemmatization, stop word removal, \etc).

\rev{The computed similarity scores yield a ranked list of UI screens and components. 
Higher-ranked screens and components in these lists are more likely to manifest  or be associated with the bug.}

\subsection{Evaluation Metrics}
\label{subsec:metrics}

We used standard retrieval metrics, widely used in prior studies~\cite{akbar2020large,lee2018bench4bl,florez2021combining,Cooper:ICSE21}, to measure the effectiveness of the studied models:

\begin{itemize}[leftmargin=*]
	\item \textbf{Mean Reciprocal Rank (MRR)}: it gives a measure of the average ranking of the \textit{first} buggy UI screen/component in the candidate list given by a model. It is calculated as: $MRR=\frac{1}{N}\sum^{N}_{i=1}\frac{1}{rank_i}$, for $N$ queries (${rank_i}$ is the rank of the first buggy UI screen/component for query $i$).
	
	\item \textbf{Mean Average Precision (\textbf{MAP})}: it gives a measure of the average ranking of \textit{all} the buggy UI screens/components for a query. It is computed as: $MAP = \frac{1}{N}\sum^{N}_{i=1}\frac{1}{BU}\sum^{BU}_{b=1}P_i(rank_b)$, where $BU$ is the set buggy UI screens/components for query~$i$, $rank_b$ is the rank of the buggy UI screen/component $b$, and $P_i(k)=\frac{buggy\_elements}{k}$ is the number of buggy UI screens/components in the top-$k$ candidates.    
	
	\item \textbf{Hits@K (\textbf{H@K})}: it is the percentage of queries for which a buggy UI screen/component is retrieved in the top-K candidates.
\end{itemize}
All metrics give a normalized score in [0, 1]---the higher the score, the higher the retrieval performance of the models.
We executed the models and the baseline approach on the constructed query sets for \rev{screen and component localization} and computed/compared the metrics between these approaches.
\looseness=-1

\subsection{Results}
\label{sec:results}

We present and discuss the effectiveness of the results of the approaches for both screen (SL) and component localization (CL). We focus our discussion on MRR since the other metrics show similar trends to the MRR results for all the models. Our replication package contains the results of all the experiments we conducted~\cite{package}.

\subsubsection{\ref{rq:screen_retr}: \textbf{Screen Localization (SL) Results}}
\label{sec:screen_retr_results}

\Cref{tab:screen_retr_results} shows the \rev{screen localization} performance of the approaches for 228 queries. The results reveal that \blip performs the highest (0.457 MRR), outperforming the second best \sbert (0.415 MRR) and the third best \lucene (0.411 MRR) with a relative improvement of 10.1\% and 11.31\% respectively. While \lucene outperforms \clip (0.381 MRR) by 7.87\%, it fails to retrieve buggy screens in 18 cases (\ie 7.89\% - not shown in the table). However, \lucene achieves a competitive H@1 to \blip (0.285). In terms of H@K, \blip outperforms the remaining models by a considerable margin. For example, it outperforms the models with a maximum relative improvement of 20.47\% H@5 (compared to \sbert). The models other than \blip achieve a similar H@5 (\clip: 0.592, \lucene: 0.575, and \sbert: 0.557).

\begin{table}[t]\centering
	\centering
	\caption{Screen localization (SL) results}
	\label{tab:screen_retr_results}
	\resizebox{\columnwidth}{!}{%
		\begin{tabular}{c|cc|ccccc}
			\toprule
			\textbf{Approach} & \textbf{MRR} & \textbf{MAP} & \textbf{H@1} & \textbf{H@2} & \textbf{H@3} & \textbf{H@4} & \textbf{H@5} \\ \hline
			\textbf{\blip}   & 0.457 & 0.443 & 0.285 & 0.447 & 0.518 & 0.592 & 0.671 \\ \hline
			\textbf{\sbert}  & 0.415 & 0.385 & 0.259 & 0.390 & 0.456 & 0.526 & 0.557 \\ \hline
			\textbf{\lucene} & 0.411 & 0.384 & 0.285 & 0.386 & 0.465 & 0.522 & 0.575 \\ \hline
			\textbf{\clip}   & 0.381 & 0.348 & 0.206 & 0.338 & 0.465 & 0.526 & 0.592 \\ \bottomrule
		\end{tabular}%
	}
\end{table}

\begin{table}[t]\centering
	\centering
	\caption{Component localization (CL) results}
	\label{tab:component_retr_results}
	\resizebox{\columnwidth}{!}{%
		\begin{tabular}{c|cc|ccccc}
			\toprule
			\textbf{Approach} & \textbf{MRR} & \textbf{MAP} & \textbf{H@1} & \textbf{H@2} & \textbf{H@3} & \textbf{H@4} & \textbf{H@5} \\ \hline
			\textbf{\sbert}  & 0.517 & 0.504 & 0.339 & 0.512 & 0.598 & 0.701 & 0.744 \\ \hline
			\textbf{\blip}   & 0.424 & 0.405 & 0.244 & 0.417 & 0.500 & 0.567 & 0.614 \\ \hline
			\textbf{\clip}   & 0.413 & 0.399 & 0.244 & 0.386 & 0.472 & 0.567 & 0.618 \\ \hline
			\textbf{\lucene} & 0.398 & 0.355 & 0.311 & 0.441 & 0.480 & 0.504 & 0.512 \\ \bottomrule
		\end{tabular}%
	}
\end{table}

The results show that \blip is the most effective model for \rev{screen localization} by a significant margin. This indicates that its rich representations, learned from images and text from other domains, can be transferred to the Buggy UI Localization problem. Also, the results imply that both sources of information (UI pixels and text) are beneficial for screen localization. Interestingly, \sbert performs second and outperforms \clip with a relative improvement of 8.9\% MRR. These results stem from the higher H@1-2 results achieved by \sbert. This is interesting and somewhat unexpected as \sbert only utilizes textual information (making it potentially less expensive to execute), while \clip uses both visual and textual information. \lucene is possibly the least expensive approach, and performs comparably to \sbert, yet it fails to retrieve buggy UI screens in 7.8\% of the cases. In line with the original \blip evaluation~\cite{li2022blip}, \blip outperforms \clip, which can be explained by the models' architecture. \blip is a model particularly designed for textual-image matching that includes a matching loss for aligning text phrases and images, learning joint representations of both sources via a contrastive loss. In contrast, \clip aims to learn representations of both textual phrases and images in the same embedding space without performing any matching.

The results also indicate that there is still room for improvement, as the best model (\blip) can suggest the buggy UI screens in the top-1 to top-3 recommendations in about 29\% to 52\% of the cases (see the H@1-3 results in \cref{tab:screen_retr_results}). A more sophisticated model is required to perform screen retrieval more effectively. Such a model should leverage both the visual information of a screen image and the textual information from the UI metadata of that screen.

\subsubsection{\ref{rq:component_retr}: \textbf{Component Localization (CL) Results}}
\label{sec:component_retr_results}

\Cref{tab:component_retr_results} shows the component localization results of all the approaches for 254 retrieval tasks. Note that although the number of OBs is 228, some OBs may have a different component corpus for retrieval, one for each buggy screen in the ground truth (\ie one OB/bug description may correspond to multiple buggy UI screens).

The results reveal that all supervised approaches perform higher (0.413+ MRR) than the baseline (\lucene), which achieves 0.398~MRR.  \sbert is the most effective of all approaches (0.517 MRR), significantly outperforming \blip, \clip, and \lucene by  21.86\%, 25.24\%, and 29.7\%,  respectively. The superiority of \sbert is consistently observed across all the metrics and it can suggest the buggy UI components in the top-1 to top-3 results in about 34\% to 60\% of the cases. 
\looseness=-1

As in screen localization, \blip slightly outperforms \clip, yet both models achieve similar performance for H@1, 4, and 5.  Although \lucene~\rev{(0.398 MRR)} achieves similar performance to \clip's (0.413 MRR) and \blip's (0.424 MRR) and higher H@1, 2, it fails to retrieve buggy UI components in 66 of 254 tasks (25.98\%).

Several observations can be derived from these results. First, the superiority of the supervised models compared to \lucene suggests that DL models are better for component localization. Second, there is still room for improving component localization: while the performance of the best model is not low, the performance is not very high either, which means that specialized models for component localization are needed. Third, the textual information present in the UI components of the screens seems to be highly effective in performing localization, as indicated by \sbert results. 
Fourth, \blip's superiority over \clip stems from their architectural differences (as discussed in \Cref{sec:screen_retr_results}).
Fifth, while it may be counter-intuitive that \sbert outperforms the multi-modal approaches, we generally observed that OBs tend to describe the buggy components using a language that is more similar to the component text observed by the user, which a language model like \sbert is specifically designed for. While \blip also leverages textual information from components, it does so based on the pixel data rather than the actual component text extracted from the UI metadata.

\subsubsection{\ref{rq:query_quality}\textbf{:} \textbf{Results by query quality and retrieval difficulty}}

\begin{figure}[t]
	\centering
	\includegraphics[scale=.42]{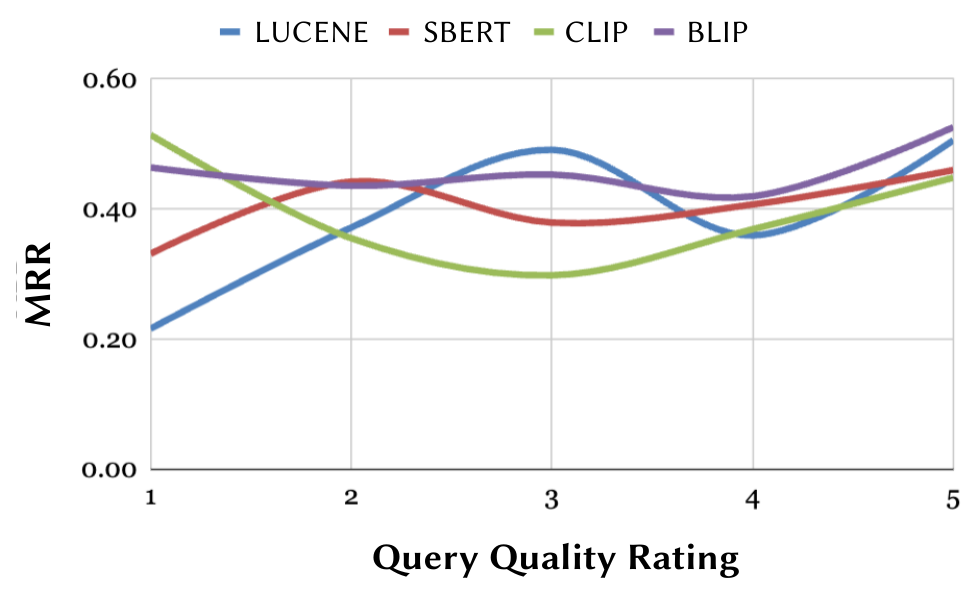}
	\caption{SL results for different query quality levels}
	\label{fig:sr_ob_rating}
\end{figure}

\textbf{\\Query Quality.} 
\Cref{fig:sr_ob_rating} shows the screen localization results (based on MRR) across different query quality ratings (from 1 to 5, 5 meaning most informative). The figure shows that while different approaches perform differently across the quality ratings, all models achieve the best performance for the most informative queries (\ie rating 5). Moreover, the performance trend is similar for all the models on the queries with quality ratings 4 and 5. Interestingly, \rev{of 18 queries for which \lucene fails to retrieve a buggy screen, eight of them have a rating of 1, and} the MRR achieved for the remaining successful cases is relatively high.

\begin{figure}[t]
	\centering
	\includegraphics[scale=.42]{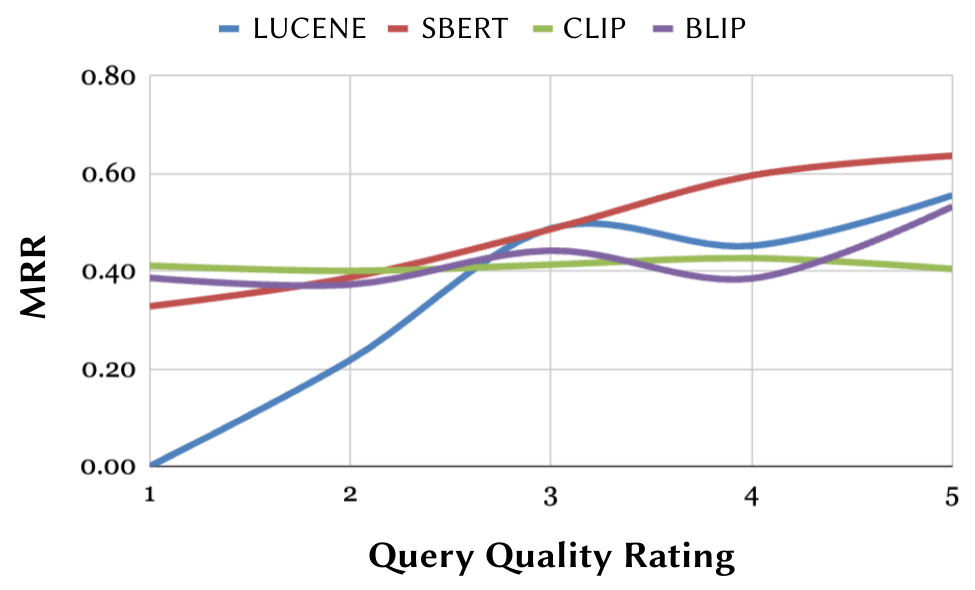}
	\caption{CL results for different query quality levels}
	\label{fig:cr_ob_rating}
\vspace{-0.5em}
\end{figure}

\Cref{fig:cr_ob_rating} shows the component localization results (based on MRR) across different query quality ratings. The figure shows a clear trend: the models tend to perform better for higher-quality queries (rating 4 \& 5) than lower-quality queries (rating 1 \& 2). As in \rev{screen localization}, of \rev{66 queries for which \lucene fails to retrieve the buggy components, most of them (43) have a rating of 1 or 2}.  Of the 17 queries with a rating of 1, \lucene fails to retrieve the \rev{UI components} for 14 queries. For the remaining 3 queries, it cannot retrieve any relevant component resulting in a 0 MRR.

For \rev{screen localization}, we found a medium-to-high positive correlation between the OB quality and the MRR results: a Spearman’s correlation of 0.41 to 0.8 across all models except CLIP.  For \rev{component localization}, we found a high correlation: Spearman’s correlation of 0.72 to 0.99 across all models. The results show that the models tend to perform better for higher-quality queries than lower-quality queries for both \rev{screen and component localization}. Our replication package contains the \# of queries per quality ratings~\cite{package}.
\looseness=-1

\begin{figure}[t]
	\centering
	\includegraphics[scale=.42]{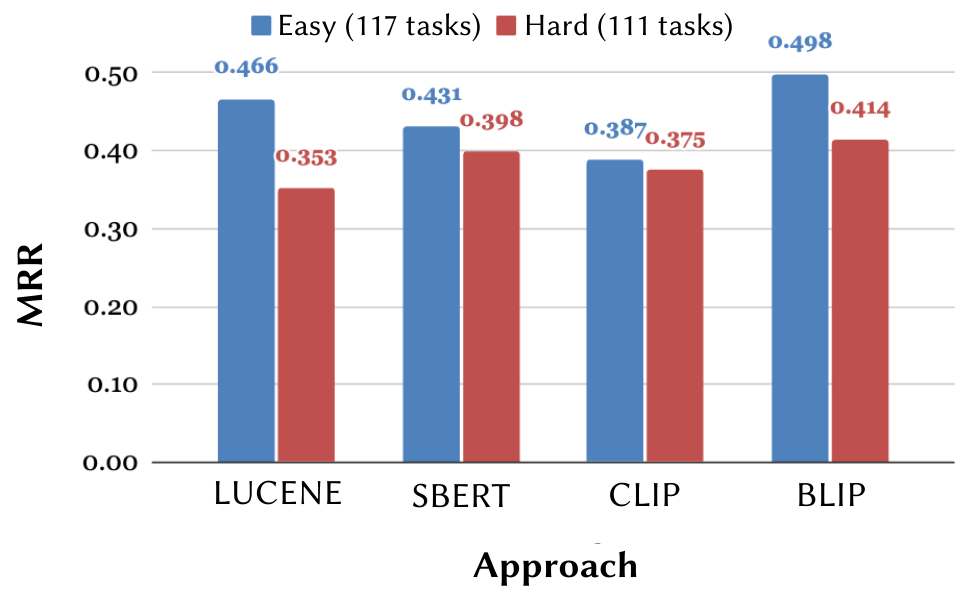}
	\caption{SL results for easy- and hard-to-retrieve tasks}
	\label{fig:sr_ob_difficulty}
\end{figure}

\begin{figure}[t]
	\centering
	\includegraphics[scale=.42]{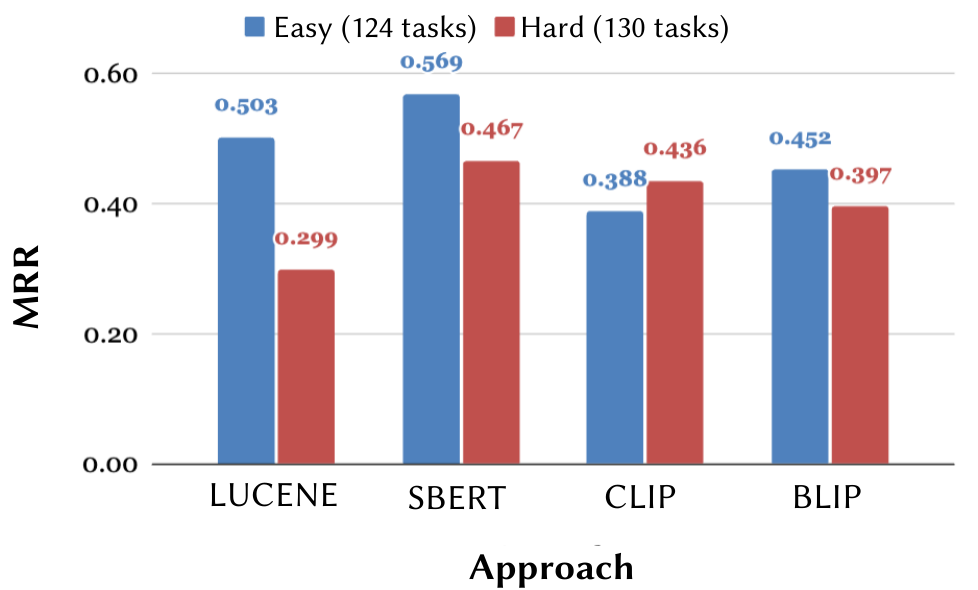}
	\caption{CL results for easy- and hard-to-retrieve tasks}
	\label{fig:cr_ob_difficulty}
	\vspace{-0.5em}
\end{figure}

\noindent\textbf{Retrieval Difficulty.}
\Cref{fig:sr_ob_difficulty,fig:cr_ob_difficulty} show the results (based on MRR) for easy- and hard-to-retrieve retrieval tasks, for \rev{screen and component localization} respectively. For SL, all models perform higher on easy-to-retrieve tasks. The same results are found for \rev{component localization}, except for \clip. The biggest performance gap is observed for \lucene (31\% for SL and 68.6\% for CL).  Of 18 failed \rev{screen localization} cases for \lucene, 3 tasks are easy and 15 tasks are hard to retrieve, and of 66 failed \rev{component localization} cases, 17 tasks are easy and 49 tasks are hard to retrieve. Regardless of the difficulty of the tasks, \blip performs highest for \rev{screen localization}, and \sbert performs highest for \rev{component localization}. The results suggest a correlation between the difficulty of retrieval by humans and the retrieval performance of the models: they tend to perform higher/lower for easier/harder cases.
\looseness=-1

\rev{The results of the different models stem from their distinct architectures, training datasets, and types. Being heavily dependent on word overlap, \lucene may fail to retrieve any UI screen/components resulting in an MRR of 0. As among the failed cases of \lucene, the majority of the queries have a lower quality level and are hard to retrieve, it exhibits the worst performance in \Cref{fig:sr_ob_rating,fig:cr_ob_rating} and the largest gaps between easy- and hard-to-retrieve cases in \Cref{fig:sr_ob_difficulty,fig:cr_ob_difficulty}. For the other three models, it is always guaranteed that no matter whether there is textual/visual similarity or not, the models will retrieve the desired UI screen/UI component in some position.
}
\looseness=-1

\subsection{Discussion}
\label{sec:results_discussion}
\rev{\textbf{Screen localization \textit{vs.} component localization}.} We found performance differences between \rev{screen localization (SL)} (0.381 - 0.457 MRR) and \rev{component localization (CL)} (0.398 - 0.517 MRR). Several factors make SL more challenging than CL. First, the corpus size is larger for SL than for \rev{CL} (25.97 screens per app vs. 17.11 components per screen on avg.). Second, SL is more abstract or general than CL as the scope of SL is broader (all screens of the application vs. all components of a screen). Additionally, OBs are generally written focusing on the component level as the user interacts with the component while reproducing the bug. Third, the quality of the OBs has an important impact on the results. For instance, \textit{``The color is unset."} is an OB with a quality rating of 2. The best SL model, \blip, identified the relevant screen for this OB in the 21st place. However, the best CL model, \sbert, identified the relevant component in the 1st place. 
\looseness=-1

\textbf{Textual \rev{\textit{vs.}} multi-modal models.} 
\rev{
For screen localization, \blip performs best on all metrics, while, for component localization, \sbert is the best performing. This distinction stems from the fundamental differences in the tasks: the retrieval corpus for component localization consists of components from the buggy screen, requiring a model that can distinguish subtle differences among components without necessarily understanding the entire screen. Additionally, when reporting bugs related to specific components, users often use the text displayed in UI components to precisely describe the issues. Therefore, text-based models like \sbert are effective at capturing the semantic meanings of these textual descriptions, making them more suitable for component localization.
\looseness=-1

However, for screen localization, multimodal models like \blip which integrates both textual and visual data, excel in understanding the full context of a screen. 
They offer detailed insight into spatial and functional interactions of various UI elements, which is crucial for localizing buggy UI screens.
Moreover, multimodal models excel not only in capturing the overall screen content but also in demonstrating strong grounding capabilities by locating specific UI components within the screen based on textual bug descriptions. 
This capability is particularly crucial for screen localization, as it allows the model to identify and differentiate screens by accurately pinpointing the relevant components when bugs are related to specific UI components.
 
}

While we found that both textual and textual-visual models achieve a reasonable performance for Buggy UI Localization, no single type of model seems to stand out. \blip and \sbert were the best-performing approaches, yet no single model was the best for both tasks. The results indicate that both types of information, textual and visual, can be leveraged for  Buggy UI Localization, yet textual data seems to be more useful for component localization, while visual data seems to be useful for screen localization.

\textbf{Design requirements for Buggy UI Localization approaches.} The results suggest that both textual and visual information alone are helpful for Buggy UI Localization. However, there is still room for improving the localization performance and specialized models may be required for this. We believe that both visual and textual information of the UI screens should be blended to build a more sophisticated model to increase localization performance. Other sources that can be explored are UI hierarchy information, which has shown promising results for command/instruction UI grounding~\cite{ li2020mapping, li2021vut}. Moreover, for a successful localization approach, we may require potentially distinct models for \rev{screen and component localization.}
\looseness=-1

Finally, while our study showed that it is feasible to leverage the pre-trained models for Buggy UI Localization, fine-tuning may be required to increase the performance of these models. However, creating or obtaining a comprehensive dataset for model fine-tuning is challenging because it should include OB descriptions of different types of bugs and wordings found in real bug reports, with corresponding ground truth data. At the same time, such a dataset should include a variety of mobile apps and should be sufficiently large for the models to effectively learn patterns from the data. Creating a global model that applies to any mobile app and bug description is challenging. Future work should explore the possibility of comparing global \textit{vs.} local models that work for specific apps, which brings an additional challenge: collecting sufficiently large ground truth data for individual apps.

\rev{\textbf{UI metadata quality assessment.}  To examine the impact of potential noisy UI metadata on the results, we assessed the quality of the three UI metadata attributes used by the textual models (\lucene and \sbert): component ID, label, and type. We focused on two factors:  attribute value presence and informativeness level. 

For value presence,  95.2\% of all the 36,572 UI components in our dataset have at least one attribute value. For 92.9\% of the components, there is a value for the component ID and/or label, which are potentially more informative than the component type. These results mean that, in 95.2\% of the cases, \lucene and \sbert leverage at least one piece of information from the UI components for retrieval.  
\looseness=-1

For the informativeness level, we qualitatively analyzed the attributes of 380 UI components (a statistically significant sample with 95\% confidence level and 5\% error margin).  
One author assessed and assigned a category (informative or non-informative) to the value of the three attributes. Another author reviewed the first author's categorization, agreeing in 91.4\% of the cases. We found that 95.8\% of components have a least one informative attribute (among the three attributes), which means that \lucene and \sbert leverage at least one informative value from the UI components for retrieval. Aggregating across the three attributes, we assessed the informativeness of 927 attribute values and found that 92\% of them are informative: 91.2\% of the component IDs, 89.8\% of the labels, and 94.7\% of the component types are informative.

We conclude that the UI metadata of our dataset is of high quality, thus giving high confidence in the study results and conclusions. 
\looseness=-1

}

\section{Improving Buggy Code Localization}
\label{sec:bug_loc_study}

To illustrate the usefulness of automated Buggy UI Localization, we conducted an additional study that investigated how the identified buggy UI screens by \blip, our best-performing \rev{screen localization} model, can improve traditional \rev{buggy code localization} approaches. We aim to answer the following research question:

\begin{enumerate}[label=\textbf{RQ$_\arabic*$:}, ref=\textbf{RQ$_\arabic*$}, itemindent=0cm,leftmargin=1cm]
	\setcounter{enumi}{3}
	\item \label{rq:bl_experiments}{\textit{Can the identified buggy UI screens by \blip lead to improved \rev{buggy code localization}?}}
	
\end{enumerate}

To answer this RQ, we adapted Mahmud \etal's approach~\cite{mahmud2023using} (\cref{subsec:bl_augementations}) as an end-to-end automated \rev{buggy code localization} technique (\cref{subsec:bl_methodology}), which retrieves potentially buggy files based on UI information from \blip's suggested buggy UI screens.
We defined different pipelines that combine Buggy UI Localization and \rev{buggy code localization} (\cref{subsec:bl_methodology}) and compared their performance with baseline techniques that do not use UI data (\cref{sec:bl_experiments}).
\looseness=-1

\subsection{UI-based Buggy Code Localization}
\label{subsec:bl_augementations}

Mahmud \etal~\cite{mahmud2023using} showed that mobile app UI interaction data improves the performance of four IR-based \rev{buggy code localizers} that rely on bug reports (\eg~\bugloc \cite{zhou2012should}).
Their approach modifies the initial ranking of potentially buggy code files produced by a \rev{buggy code localizer} for a given bug report, by \textbf{boosting} relevant files and/or \textbf{filtering} out irrelevant files, or by performing \textbf{query reformulation}.
These operations (\aka augmentations) use information extracted from the UI screen that shows the reported bug and the preceding 1-3 screens in a bug reproduction trace.

\rev{The information extracted from UI screens is \textit{UI terms} (\eg activity and window names) which are matched against code file names to produce a set of \textit{UI-related files}.} The UI terms and UI-related files are used by two augmentation methods: (1) \rev{\textbf{Reformulating queries} via query expansion, which appends UI terms to bug reports, or via query replacement, which uses UI terms as the query; and (2) \textbf{File re-ranking} by filtering, boosting, or combining filtering and boosting. Filtering removes files that do not match UI-related files (\eg classes that directly interact with the UI) from the file corpus. Boosting elevates the ranking of files in the corpus that match UI-related files during the search.}
\looseness=-1

\looseness=-1

Mahmud \etal employed four main configuration parameters to integrate UI information into the IR bug localizers: (1) the number of UI screens in a reproduction trace (\ie the buggy screen and the preceding 1-3 screens);
(2) five types of UI information sources (\eg UI screen, UI components, and exercised UI components);
(3) query reformulation strategies; and (4) re-ranking strategies. In total, 657 configurations were defined and evaluated for each bug localizer.

\subsection{Integrating Buggy UI and Code Localization}
\label{subsec:bl_methodology}

Mahmud \etal's approach~\cite{mahmud2023using} requires as input a trace of the \rev{UI screens and UI components} that the user interacted with to reproduce a reported bug.
The trace and the buggy UI screen in the trace are meant to be manually collected/identified by the developer. Mahmud \etal's approach then uses the metadata information from the buggy screen and the 1-3 prior screens/components in the reproduction scenario as input to their augmentation approaches.
\looseness=-1

Our goal is to eliminate the manual effort of Mahmud \etal's approach and define an effective and fully automated end-to-end pipeline of \rev{buggy code localization} using the buggy UI screens recommended by a UI screen localizer. To that end, we adapted their approach by using our best screen localization approach (\ie~\blip) to automatically suggest the top 3-4 buggy UI screens as the only information needed by the \rev{buggy code localization} pipeline.  
\looseness=-1

As such, we defined an approach that integrates both the screen localization and \rev{buggy code localization} pipelines since the ultimate goal is to produce a ranked list of potentially buggy code files for a given bug report. The challenge in defining this combined approach is that a bug report can contain multiple OB descriptions. If we execute \blip on each OB description, it would produce multiple lists of potentially buggy UI screens. Therefore, this challenge is rooted in deciding which buggy screens should be given as input to the localization pipeline, to produce a single ranking of buggy files.  

To address this problem, we considered two options: (1) produce and provide a single ranking of UI screens for the bug report, or (2) provide each ranking of UI screens (for each OB description in the bug report) to the \rev{buggy code localization} pipeline, to produce multiple code file rankings, and then combine these rankings into a final code file ranking. For option \#1, we explored two strategies, namely: (i) \rev{\textbf{\concatob}}, which concatenates the OB descriptions in a bug report and uses the resulting query as input to \blip, and (ii) \rev{\textbf{\firstob}}, which selects only the first OB description found in the bug report as a query to \blip. \rev{When executed,} these two strategies 
produce a single UI screen ranking, which can be used by the \rev{buggy code localization} pipeline to suggest a single code file ranking for the bug report. As for option \#2, to produce a single code file ranking, \rev{we defined a strategy called \textbf{\indivob}}, which first averages the similarity scores of each code file found in all the buggy file rankings to produce a single similarity score for the file. Then, these combined similarities, for all the files in the rankings, are used to produce a final code file ranking (\ie sorting by these similarities).
\looseness=-1

\subsection{Approach Execution, Dataset, and Metrics}
\label{sec:bl_experiments}

We selected the two best IR-based \rev{buggy code localization} techniques from Mahmud \etal's study~\cite{mahmud2023using}, namely \lucene~\cite{Hatcher2004} (adapted for \rev{buggy code localization}) and \bugloc~\cite{zhou2012should}, and executed them in our combined pipeline for \rev{buggy code localization}. We tested all 70 feasible configurations of different augmentation methods and UI information kinds defined in the prior work. We also experimented with providing the top 3 and 4 buggy UI screens suggested by \blip, based on the best number of screens found in the prior study for \lucene (4 screens) and \bugloc (3 screens).
\looseness=-1

We executed the three combined pipelines defined above, namely \textbf{\concatob},\textbf{ \firstob}, and \textbf{\indivob}, using both buggy code localizers. However, we could not execute \indivob with \bugloc because the tool provided by the original authors~\cite{zhou2012should} does not provide the code file rankings needed by \indivob. The pipelines were executed on 79 of the 80 bug reports from the \rev{buggy code localization} benchmark provided by Mahmud \etal~\cite{mahmud2023using}. As mentioned in \cref{subsec:bug_report_selection}, we excluded one bug report because we could not reproduce the bug. The full set of 87 bug reports was not used in the Buggy UI Localization due to the absence of Java files as ground truth for \rev{buggy code localization}. 

The performance of the combined pipelines, using all possible configurations and IR bug localizers, was measured and compared using Hits@k and its relative improvement (RI), in line with the methodology followed by the prior work~\cite{mahmud2023using}. We used as baselines the original IR bug localizers, without using any UI information. 
Note that 4 screen experiments used 77 bug reports as 2 bug reports have only 3 screens in the \rev{screen localization (SL)} corpus.

\rev{
	\begin{table}[]
		\centering
		\caption{\rev{Buggy Code Loc. Performance via Buggy UI Loc.}}
		\label{tab:bl_results}
		\resizebox{\columnwidth}{!}{%
			\begin{tabular}{c|c|c|cc|c|c}
				\toprule
				\textbf{Bug Localizer} &
				\textbf{Approach} &
				\textbf{\begin{tabular}[c]{@{}c@{}}\rev{\#}\\ \rev{Scrns}\end{tabular}} &
				\textbf{H@5} &
				\textbf{H@10} &
				\textbf{\begin{tabular}[c]{@{}c@{}}RI of\\ H@10\end{tabular}} &
				\textbf{\begin{tabular}[c]{@{}c@{}}\#Bug\\ Top10\end{tabular}} \\ \hline
				\multirow{4}{*}{\textbf{\lucene}}     & Baseline       & \rev{4} & 0.74 & 0.79 & -     & 61 \\ \cline{2-7} 
				& \concatob     & \rev{4} & \rev{0.75} & \rev{0.88} & \rev{11.49\%} & \rev{68} \\ \cline{2-7} 
				& \firstob      & \rev{4} & \rev{0.74} & \rev{0.88} & \rev{11.49\%} & \rev{68} \\ \cline{2-7} 
				& \indivob & \rev{4} & 0.77 & 0.87 & 9.85\%  & 67 \\ \hline
				\multirow{3}{*}{\textbf{\bugloc}} & Baseline       & \rev{3} & \rev{0.59} & \rev{0.71} & -     & \rev{56} \\ \cline{2-7} 
				& \concatob     & \rev{3} & \rev{0.72} & \rev{0.79} & \rev{10.72\%} & \rev{62} \\ \cline{2-7} 
				& \firstob      & \rev{3} & \rev{0.61} & \rev{0.80} & \rev{12.41\%} & \rev{63} \\ \bottomrule
			\end{tabular}%
		}
	\end{table}
\vspace{-1em}
}

\subsection{Results}
\label{sec:bl_results}

\Cref{tab:bl_results} shows the \rev{buggy code localization} results for both IR bug localizers and the best configurations we obtained among all configurations. These results are obtained when \blip suggests the \rev{top 3 and 4 buggy screens}. 
Complete results of all configurations and experiments are provided in our replication package~\cite{package}.
\looseness=-1

For each pipeline, IR bug localizer, and number of buggy screens recommended by \blip, we consistently found that the best configuration (\ie the highest H@10 improvement compared to the baselines) \rev{includes filtering with UI  Components (SC) and boosting with UI Screens (GS). Additionally, the best configuration includes query expansion with GS when \concatob is used with \lucene and query expansion with SC when \firstob is used with  \bugloc. As in the prior work~\cite{mahmud2023using}, we obtained the best results with 4 screens for \lucene and 3 screens for \bugloc.}

\Cref{tab:bl_results} reveals that all the combined pipelines for \rev{buggy code localization} lead to a performance improvement compared to the baselines, \rev{by 9.85\% to 12.41\% H@10. When using \lucene, the \concatob and \firstob pipelines achieve the best improvement: 11.49\% for H@10, which translates into retrieving the buggy code files in the top-10 results for 7 more bug reports, compared to the baseline. When using  \bugloc, \firstob pipeline results in improving the baseline by 12.41\% H@10 (\ie 7 more successful retrieval tasks).} 
\looseness=-1

We compare our results from \cref{tab:bl_results} with the results achieved by the best configurations obtained for \lucene and \bugloc by Mahmud \etal~\cite{mahmud2023using}, since those results represent a perfect identification of the buggy UI screen, along with the reproduction scenario. However, we must cautiously compare these results since the bug reports used in both studies are not exactly the same. Not surprisingly, the performance of the manual \rev{buggy code localization} approach by Mahmud \etal~\cite{mahmud2023using} is slightly higher than the performance of our best configurations \rev{(0.9 vs 0.88 H@10 for \lucene, and 0.84 vs 0.80 H@10 for \bugloc).}  This difference is acceptable, considering that we propose a fully automated way of localizing buggy code files via Buggy UI Localization, which still outperforms baseline localizers, while the prior work requires manual effort in collecting reproduction traces and the buggy screen.

Given the results, we conclude that Buggy UI Localization can be useful to improve the performance of UI-based \rev{buggy code localization} in a fully automated end-to-end way.

\section{Threats to Validity}
\label{sec:threats}

\textbf{Construct Validity.} 
There may be subjectivity introduced in the dataset construction when identifying the OB descriptions in the bug reports, their quality rating, retrieval difficulty levels, and the ground truth \rev{buggy UI screens and components}. We mitigated this threat by adopting a rigorous methodology to label and curate the data during joint sessions of bug understanding, replication, and analysis among four authors, reaching consensus in all cases.

\textbf{Internal Validity.} The selection of models affects the internal validity of our results/conclusions. To mitigate this, we covered both uni-modal (\sbert) and multi-modal (\clip \& \blip) DL models, and a unsupervised textual technique (\lucene) as baseline for Buggy UI Localization. 
For buggy code localization study, we conducted various experiments with all feasible configurations on two localizers (\lucene and \bugloc) to obtain the best-performing configuration. 
\rev{Another threat concerns the methodology we used to collect UI corpus data, 
which may have led to incomplete coverage of UI screens for an app. While this can have an impact on the study results, we used the same screen corpus to evaluate all models, thus ensuring a fair evaluation.}

\textbf{External Validity.} The conclusions of our study may not generalize to other retrieval models, bug descriptions, and apps. To improve the generalization, we selected different types of models and built a real dataset containing a variety of bug types, and apps that implement different UIs for multiple domains.
\looseness=-1

\section{Related Work}
\label{sec:related_work}

\textbf{Mobile App Bug Report Management.} 
\rev{Recent research \cite{feng2022gifdroid, song_toward_2022, Fazzini:TSE22, zhang2023automatically, feng2023read}} has explored the use of mobile app bug reports to automate various bug report management tasks.
Researchers~\cite{Zhao2019,zhang2023automatically} have proposed approaches to reproduce Android bugs or generate test cases based on bug reports. 
Our Buggy UI Localization approach that identifies the buggy screens/components can help these approaches to generate assertions that validate the reported bugs.
Song \etal~\cite{song_toward_2022} proposed a  chatbot to help users report Android bugs via visual guidance and quality verification. This chatbot can benefit from a Buggy UI Localization approach by accurately assessing how bug  descriptions corresponds to UI screens/components.
Despite the growing body of research on automating bug report management tasks such as \rev{bug reporting~\cite{song_toward_2022,Fazzini:TSE22}, reproduction~\cite{Zhao2019,feng2022gifdroid,zhang2023automatically,Zhao2019,zhao2022recdroid+,Bernal:TSE22,Bernal:ICSE20}, localization~\cite{akbar2020large,
Wong2014,Kochhar2014,florez2021combining,chaparro2019using,Chaparro2017-3,chaparro2019reformulating}, and others~\cite{feng2023read,Zhou2012a,yan2024semantic,Cooper:ICSE21}, 
prior work has not explored how to automatically localize buggy UIs as we do.}

\rev{
\textbf{Static analysis of mobile app UIs.} Researchers have proposed techniques/tools to statically analysis mobile app UIs~(\eg \, \frontmattert~\cite{kuznetsov2021frontmatter,kuznetsov2021all}, \gator~\cite{yang2018static}, \backstage~\cite{avdiienko2017detecting}, \goalexplorer~\cite{lai2019goal}, and others~\cite{li2023crowdsourced,guo2020improving}). 
One main advantage of these approaches, over dynamic analysis, is their ability to cover a large number of UI screens for an application.
Our future work will explore the use of these techniques to assess how UI screen coverage can impact the performance of the studied models. Static analyzers also provide features that can assist buggy code localization. For example, \frontmattert~\cite{kuznetsov2021frontmatter,kuznetsov2021all} identifies which Android APIs can be triggered by an interaction with a UI component, which may help identify buggy code elements. While the main problem we address in this paper is buggy UI localization, our future work will explore how static analyzers and UI localizers can be integrated to better localize buggy code based on a bug description. The main limitations of static analyzers, which prevented us from using them for data collection, include: (1) they may fail to capture server-side content loaded only at runtime, resulting in potentially unrealistic UI screens (2) they do not provide UI screenshots, needed to evaluate the studied multi-modal models, and (3) potential imprecision of app behavior captured by these tools as they may not provide UI screens displaying certain bugs (\eg incorrect output and navigation issues).
}
\looseness=-1

\textbf{UI Representation Learning and Applications.} UI representation learning aims to represent UI elements or text via embeddings~\cite{li2021screen2vec,he2021actionbert, bai2021uibert,li2021vut} for downstream tasks such \rev{as image captioning~\cite{Wang:UIST21, moran2022empirical, chen2022towards} and UI component labeling  \cite{chen2020unblind, li2020widget, chen2022towards}.
} One application of UI representation learning is mapping (\aka grounding) textual instructions to UI action/elements~\cite{pasupat2018mapping, li2020mapping, xu2021grounding}. Pasupat \etal~\cite{pasupat2018mapping} evaluated three  models to ground natural language commands to web elements. Li \etal~\cite{li2020mapping}  \rev{utilized transformers models for this task}, based on three synthetic training datasets.  
There are key differences that make it hard to adapt those models to our problem. For example, Li \etal's approach~\cite{li2020mapping}  requires a sequence of screens where the instructions are performed, and then locating the corresponding UI component for each instruction. In contrast, our work identifies the buggy UI screens and components without any prior information about  relevant screens. Furthermore, our study deals with bug descriptions, whose language is more complex than that of UI instructions~\cite{Chaparro2017-2}.

\section{Conclusions}
\label{sec:conclusions}

This paper reports the results of the first empirical study that investigates the effectiveness of textual/visual neural models for automatically localizing \rev{buggy UI screens and UI components} based on the bug descriptions of mobile apps. We evaluated approaches for screen and component localization, using a real-world dataset of manually-curated bug descriptions and ground truth UI data.

The study reveals that the best-performing approaches can localize correct \rev{buggy UI screens and components} in the top-3 recommendations for 52\% and 60\% of the bug descriptions. We found that the models tend to perform better for the bug descriptions which are easier to retrieve even for humans and which have a higher quality. %
We also illustrate that Buggy UI Localization can be useful to automate and improve traditional buggy code localizers.

\section{Data Availability}
For verifiability and reproducibility, we have made all the study artifacts (including the dataset, source code, and documentation) publicly available~\cite{package}.

\section*{Acknowledgments}
This work is supported by the U.S. NSF  CCF grants 1955853, 2343057, and 2007246. Any opinions, findings, and conclusions expressed herein are of the authors and do not necessarily reflect the sponsors'. 
\looseness=-1

\balance
\bibliographystyle{ACM-Reference-Format}
\bibliography{references}

\end{document}